\documentclass[12pt]{article}
\usepackage{ amsmath, amsthm, amssymb } 
\usepackage{adjustbox}

\usepackage{color} 
\usepackage{cite}

\newcommand{\be} { \begin{equation} } 
\newcommand{\ee} { \end{equation} } 
\newcommand{\nn}{\nonumber \\}

\newcommand{\Tad}{{\rm Tad}} 
\newcommand{\Bub}{{\rm Bub}} 
 
\newcommand{\Tri}{{\rm Tri}} 

\newcommand{\D}{\mathfrak{D}}
\newcommand{\I}{\mathcal{I}}

\newcommand{\N}{\mathbb{N}}

\newcommand{\Cut}{ \text{Cut}}
\newcommand{\kk}{\bar{k}}

\newcommand{\e}{\epsilon}

\DeclareMathOperator{\EK}{K}

\begin{document} 
\setlength{\unitlength}{1.3cm} 
%%%%%%%%%%%%%%%%%%%%%%%%%%%%%%%%%%%%%%%%%%%%%%%%%%%%%%%%%%%%%%%%%%%%%%% 
\begin{titlepage}
\vspace*{-1cm}
\begin{flushright}
TTP16-046
\end{flushright}                                
\vskip 3.5cm
\begin{center}
\boldmath
 
{\Large\bf On the maximal cut of Feynman integrals and the solution of 
their differential equations\\[3mm] }
\unboldmath
\vskip 1.cm
{\large Amedeo Primo}$^{a,b,}$
\footnote{{\tt e-mail: amedeo.primo@pd.infn.it}} and
{\large Lorenzo Tancredi}$^{c,}$
\footnote{{\tt e-mail: lorenzo.tancredi@kit.edu}} 
\vskip .7cm
{\it $^a$ Dipartimento di Fisica ed Astronomia, Universit\`a di Padova, Via Marzolo 8, 35131 Padova, Italy } \\
{\it $^b$ INFN, Sezione di Padova, Via Marzolo 8, 35131 Padova, Italy} \\
{\it $^c$ Institute for Theoretical Particle Physics, KIT, 76128 Karlsruhe, 
Germany } 
\end{center}
\vskip 2.6cm

\begin{abstract}
The standard procedure for computing scalar multi-loop Feynman integrals 
consists in reducing them to a basis of so-called master integrals, derive 
differential equations in the external invariants satisfied by the latter and, finally, 
try to solve them as a Laurent series in $\epsilon = (4-d)/2$, 
where $d$ are the space-time dimensions. The differential equations are, in general,
coupled and can be solved using Euler's variation of constants, provided that
a set of homogeneous solutions is known. Given an arbitrary differential equation
of order higher than one, there exists no general method for finding its 
homogeneous solutions.
In this paper we show that the maximal cut of the integrals under consideration
provides one set of homogeneous solutions, simplifying substantially 
the solution of the differential equations.

\end{abstract}
\vfill
\end{titlepage}                                                                
\newpage

%%%%%%%%%%%%%%%%%%%%%%%%%%%%%%%%%%%%%%%%%%%%%%%%%%%%%%%%%%%%%%%%%%%%%%% 
\section{Introduction} \label{sec:intro} \setcounter{equation}{0} 
\numberwithin{equation}{section}
The method of differential equations~\cite{Kotikov:1990kg,Bern:1993kr,Remiddi:1997ny,Gehrmann:1999as} 
is undoubtedly one of the most powerful
techniques for the calculation of multi-loop Feynman integrals.
The latter is based on the possibility of reducing a family of Feynman integrals
to a small subset of basic integrals, the so-called master integrals (MIs), through
the repeated use of integration by parts 
identities (IBPs)~\cite{Tkachov:1981wb,Chetyrkin:1981qh,Laporta:2001dd}.
The IBPs themselves can then be used to show that the MIs fulfill
systems of linear coupled differential equations in the external invariants,
whose solution is usually much simpler than a direct integration over the loop
momenta. For a review see~\cite{Argeri:2007up,Henn:2014qga}.

We are normally interested in computing the master integrals as Laurent series
in $\epsilon = (4-d)/2$. 
To this aim, there are two fundamental steps which must be achieved.
First of all, given a sector with $N$ coupled master integrals, one needs a way to
determine the \textsl{minimum number} of coupled differential equations in the
limit $d \to 4$. This is important since it determines the effective degree of the 
differential equation that has to be iteratively solved. If all master integrals decouple (or if the
homogeneous system takes a triangular form), we
are left effectively with a series of first order differential equations in $d=4$, which can be 
solved by quadrature. If, instead, some of the integrals remain coupled, one
has to solve a higher order differential equation; naively one expects 
that the higher is the rank,  the more involved the mathematical structure of the solutions will be.  
A way towards a systematical determination of the minimum degree of the coupled equations
has been suggested in~\cite{Remiddi:2013joa,Tancredi:2015pta}, where it was shown that 
the information useful to decouple some of the differential equations in $d=4$
can be read off from the integration by parts identities close to $d=2\,n$, with
$n \in \N$. 

The choice of master integrals is, of course, arbitrary and more recently it was
shown that by properly selecting the basis of integrals one can simplify the form 
of the differential equations substantially, bringing them
to a so-called canonical form~\cite{Henn:2013pwa}. One of the fundamental properties
of a canonical form is that the \textsl{homogeneous part} of the system of 
differential equations becomes trivial in the limit $\epsilon \to 0$
$$\partial_x\, \vec{m} = \mathcal{O}(\epsilon),$$
where $\vec{m}$ is the vector of master integrals and $x$ is a generic 
external invariant. This implies that the homogeneous solution
for $\epsilon=0$ is a constant. 
Different algorithms 
have been proposed for the construction of a canonical basis, 
starting from the properties of the system of differential equations.
Currently, they are all limited to special situations, such as a linear dependence on $\epsilon$ \cite{Argeri:2014qva} or the dependence on a single kinematic 
variable~\cite{Lee:2014ioa,Ablinger:2015tua,Lee:2016lvq}.
Steps towards algorithms valid also in the case of several variables 
have been made in~\cite{Gehrmann:2014bfa} and more recently and thoroughly in~\cite{Meyer:2016slj}.
Both papers make use of Ans\"{a}tze for the linear combinations of master integrals
which fulfill canonical differential equations. While such approaches are often very useful,
their applicability remains for now limited to those cases where all square roots can be removed and
the alphabet can be completely rationalized. As it is well known, this is not always possible
even when a canonical basis is known to exist, for a recent example see~\cite{Bonciani:2016qxi}.
On the other hand, in~\cite{Henn:2013pwa} it was suggested that 
one of the crucial criteria for selecting candidate integrals 
which fulfill canonical differential equations is the study of their leading singularity, 
which can be done  by inspecting their generalized cuts~\cite{Henn:2014qga}.
While a complete understanding of the concept of leading singularity and
of a canonical basis are available only in the case of integrals that evaluate
to multiple-polylogarithms and generalizations thereof\footnote{We 
refer here to Feynman integrals whose solution can be written as iterated integrals over
d-log forms of arbitrarily complicated (i.e. not necessarily linear) arguments.}~\cite{ArkaniHamed:2010gh}, 
it has been suggested that the study of the maximal cut 
should be the starting point to extend these considerations also to more 
complicated cases\cite{Bonciani:2016qxi}.
A crucial observation
is that the maximal cut provides by construction exactly a solution 
of the homogeneous system of differential equations.
The connection between unitarity cuts and differential equations 
is not new. 
In~\cite{Laporta:2004rb}, for example,
the second order differential equation satisfied by the two-loop massive sunrise
graph was solved by inferring the homogeneous solution from the calculation of 
the imaginary part of the graph.
More in general, this connection has been largely exploited in the so-called
reverse unitarity~\cite{Anastasiou:2002yz} method, while a new way of solving IBPs
using the information coming from the unitarity cuts has been proposed 
in~\cite{Larsen:2015ped}.
Finally, similar conclusions to the ones drawn in this paper 
have been already exploited in the context of the DRA method
based on dimensional recurrence relations~\cite{Lee:2012te}.

A way to algorithmically find a homogeneous solution of a coupled
system of equations is indeed very desirable.
In fact, once we are confident to have reduced the degree of the differential equation
to the minimum possible, the problem of finding an integral representation for the
solution relies on the ability to solve its homogeneous part. 
Once a homogeneous solution is known,
one can use Euler's method of variation of constants in order to build up the
inhomogeneous solution.
The goal of this paper is to show that the maximal cut can be used in many non-trivial
examples as a powerful tool to compute the homogeneous solutions,
in particular when the latter can be written in terms of complete elliptic integrals.
One should recall, in fact, that given a coupled system of $N$ differential equations,
a complete solution requires finding $N$ linearly independent solutions.
In the case of elliptic integrals, one is usually left with systems of two coupled equations, which require therefore finding two independent solutions.
While the maximal cut provides only one solution,
once this is known, the second can be obtained 
by simply exploiting the properties of the complete elliptic integrals.

The rest of the paper is organized as follows. In Section~\ref{sec:general}
we set up the notation, summarize the general idea and give a prescription
to compute the maximal cut of integrals with squared propagators.
In Section~\ref{sec:oneloop} we apply the method to a simple one-loop example and
show explicitly the one-to-one relation between the maximal cut and the solution of its homogeneous differential equation. This 
relation is true for any values of $d$ and works of course both ways, which implies that
it also provides us with a 
powerful tool to compute the maximal cut of any one-loop integral in $d$ dimensions:
this can be done simply by solving its homogeneous differential equations.
In Section~\ref{sec:twoloop} we then move to more interesting two loop examples
which evaluate to integrals over elliptic integrals.
Finally we conclude in Section~\ref{sec:conclusions}.

%%%%%%%%%%%%%%%%%%%%%%%%%%%%%%%%%%%%%%%%%%%%%%%%%%%%%%%%%%%%%%%%%%%%%%%%%% 

\section{Maximal cut and differential equations}
\label{sec:general} \setcounter{equation}{0} 
\numberwithin{equation}{section} 

Let us consider a family of $l$-loop Feynman integrals with $r$ different propagators and
$s$ irreducible scalar products, 
\begin{equation}
\I(d;x;a_1,...,a_r;b_1,...,b_s) = \int \prod_{j=1}^l \D^d k_j\, \frac{S_1^{b_1}\,...,S_s^{b_s}}{D_1^{a_1}\,...\,D_r^{a_r}}\,. \label{eq:intfam}
\end{equation}
The integrals are functions of the dimensional regularization parameter $d$
and of some external
invariants which we call collectively $x$. First of all, we should define what 
we mean with cutting an integral and, in particular, with its maximal cut.
It is well known that  
unitarity cuts can be used to study the discontinuity of Feynman graphs with respect to
a given Mandelstam invariant. The original integral can then be reconstructed from
its discontinuities by using dispersion 
relations~\cite{Cutkosky:1960sp,Veltman:1963th,Remiddi:1981hn}.
For more recent developments in this directions see for example~\cite{Abreu:2014cla, Abreu:2015zaa}.
Here we will not enter into this fascinating and rich subject.
The only thing we will need is an operative definition of maximal cut for an
integral of the form~\eqref{eq:intfam}.

Cutting a propagator in a Feynman diagram means, loosely speaking, to 
force the particle propagating through it to be on-shell; mathematically,
we can achieve this relatively easily if the propagator is raised to power one.
In this case, we can simply substitute the propagator with a Dirac $\delta$-function
which forces the momentum of the propagator on-shell by imposing a constraint
on the components of the loop momenta. We define the maximal cut of a diagram 
as the simultaneous cut of all its propagators and we will always assume such multiple cut to exist. In fact, if the set of on-shell conditions for a given diagram does not admit any solution, the diagram can be proven to be reducible, i.e. it can be trivially expressed as a linear combination of diagrams with a reduced number of propagators\cite{Mastrolia:2012an}. In the rest of our discussion will focus on diagrams corresponding to irreducible topologies, whose maximal cut is therefore a well defined object.
As we will show in the following, we found it convenient to compute the 
maximal cut of two-loop graphs by first localizing one of the one-loop sub-diagrams, 
and then integrating directly the remaining $\delta$-functions obtained cutting the
second loop. By appropriately choosing which sub-loop to cut first, the computation can be
substantially simplified. Moreover
this procedure can easily be generalized for higher loops.
We note here that an alternative way of performing the maximal cut is going through
Baikov's representation for the loop integrals~\cite{Baikov:1996iu}, as explained
for example in~\cite{Larsen:2015ped}.

\subsection*{Cutting squared propagators}
It is useful to define what we mean with cutting a squared propagator.
Indeed, at one loop one is always left with at most one master integral 
per topology, which can always be chosen with propagators raised at most to 
power one. On the other hand, in a generic multi-loop calculation, more integrals 
can remain independent and we might have to consider also
integrals with squared internal propagators. 
In what follows, we will use two operative prescriptions to cut a squared propagator, which produce
equivalent results. We will use these prescriptions to compute the maximal cut but, of course,
they can be applied for the computation of any other cut.

\begin{enumerate}
\item The first prescription is based on integration by parts identities.
Given any family of Feynman integrals, we can use IBPs to write any integral  with
squared propagators in terms of similar integrals with propagators raised at most to unit
powers and, possibly, scalar products at the numerator.
Let us take such an integral $\I_{\rm dot}(d;x)$ and write
\begin{align}
\I_{\rm dot}(d;x) = \sum_{j=1}^N\, c_j(d;x)\,m_{j}(d;x) + {\rm subtopologies}\,, \label{eq:cutsqr1}
\end{align}
where $m_{j}(d;x)$ are the $n$ linearly independent master integrals which do not
contain any squared propagator. The $c_j(d;x)$ are instead rational functions
in the dimensions $d$ and in the Mandelstam invariants $x$.

If we apply now a maximal cut on~\eqref{eq:cutsqr1} we get 
\begin{align}
\Cut(\I_{\rm dot}(d;x) ) = \sum_{j=1}^N\, c_j(d;x)\, \Cut( m_{j}(d;x) )\,, \label{eq:defcutsqr1}
\end{align}
where we used the fact that the maximal cut of the subtopologies is identically zero.

\item An alternative prescription to cut a squared propagator is as follows.
Let us take an integral defined as
\begin{equation}
\I_{\rm dot}(d;m_1^2,...,m_s^2,...,m_r^2,x) = \int \prod_{j=1}^l\, \D^d\, k_j\, 
\frac{1}{D_1(m_1^2) ... D_s^2(m_s^2)... D_r(m_r^2)}\,, \label{eq:cutsqr2}
\end{equation}
where the propagators $D_k$ could in general depend on some masses $m_k$, some or all
of which could
of course also be zero. 
We are integrating over $l$-loops and assuming that the integral has $r$ propagators;
the $s$-th propagator is squared.
In order to compute the maximal cut of this integral, let us consider the associated
integral
\begin{equation}
\I(d;m_1^2,...,\widetilde{m}_s^2,...,m_r^2,x) = \int \prod_{j=1}^l\, \D^d\, k_j\, \frac{1}{D_1(m_1^2)... D_s(\widetilde{m}_s^2) ... D_r(m_r^2)}\,, \label{eq:cutsqr3}
\end{equation}
where all propagators have unit power and 
we have modified the mass $m_s^2 \to \widetilde{m}_s^2$ in the propagator $D_s$,
such that $\widetilde{m}_s^2$ is different from any other mass in the remaining propagators.
It is clear then that
\begin{align}
\I_{\rm dot}(d;m_1^2,...,m_s^2,...,m_r^2,x) = 
\lim_{\widetilde{m}_s^2 \to m_s^2}\; \frac{\partial}{\partial\, \widetilde{m}_s^2}
\I(d;m_1^2,...,\widetilde{m}_s^2,...,m_r^2,x)\,,
\end{align}
and therefore
\begin{align}
\Cut( \I_{\rm dot}(d;m_1^2,...,m_s^2,...,m_r^2,x) )= 
\lim_{\widetilde{m}_s^2 \to m_s^2}\; \frac{\partial}{\partial\, \widetilde{m}_s^2}
\Cut( \I(d;m_1^2,...,\widetilde{m}_s^2,...,m_r^2,x) )\,. \label{eq:defcutsqr2}
\end{align}
If the limit is smooth, this provides us with a second operative prescription for computing the
maximal cut of a graph with a squared propagator.

\end{enumerate}

\subsection*{Differential equations and the maximal cut}
Let us switch now to the connection between differential equations and the
maximal cut defined above.
Let us consider again a family of Feynman integrals like in Eq.~\eqref{eq:intfam},
which is reduced to $N$ independent master integrals
$m_i(\epsilon;x),...,m_N(\epsilon;x)$. 
The master integrals satisfy a system of $N$ coupled differential equations in the external
invariants
\begin{align}
\partial_x \, m_i(d; x) = H_{ij}(d; x)\,m_{j}(d; x) 
+ G_i(d; x),
\label{eq:gendeq}
\end{align}
where the $H_{ij}(d;x)$ are the coefficients of the
homogeneous system, while the $G_i(d;x)$ contain the dependence
on the sub-topologies, which are simpler graphs with fewer propagators.
Imagine now to perform a maximal cut on the master integrals $m_i(d;x)$,
which corresponds to cutting all $r$ propagators. As we discussed above,
we only consider irreducible cases where all $r$ propagators can be
simultaneously cut.
Clearly, if we imagine to apply our cutting procedure on Eq.~\eqref{eq:gendeq}
we will be left with
\begin{align}
\partial_x \, \Cut(m_i(d; x)) = H_{ij}(d; x)\,\Cut(m_{j}(d; x))\,,
\label{eq:gendeqcut}
\end{align}
where we use the notation $\Cut(m_i(d;x))$ for the maximal cut of the
integral under consideration and we have obviously 
\begin{equation}
\Cut(G_i(d;x)) = 0\,. \label{eq:cutsubtopos}
\end{equation}
Eq.~\eqref{eq:cutsubtopos} should be obvious, since all integrals contained in $G_i(d;x)$
contain fewer propagators than the master integrals $m_i(d;x)$ and, consequently, applying the
same cut on the latter must produce zero. 
We should stress here that, in general, the maximal cut exists only for
complex values of the kinematical invariants. 
Eqs.~(\ref{eq:gendeqcut},~\ref{eq:cutsubtopos})
are the central observations on which this paper is based. 
Similar conclusions in the context of the DRA method have been drawn in~\cite{Lee:2012te}.
What they tell us is that the maximal cut of the master integrals $m_i(d;x)$
must satisfy the homogeneous part of the system~\eqref{eq:gendeq}.
A remark here is in order. It is clear that if, for any reason, the maximal
cut of any of the master integrals
$m_i(d;x)$ turns out to be zero for a given value of the dimensions $d$, this 
implies that the master under consideration must decouple from the
system~\eqref{eq:gendeqcut} for this value of $d$. 
Indeed, while the trivial solution $m_i(d;x) = 0$ is always a solution
of the homogeneous system, the latter will in general admit other non-trivial solutions,
which cannot be computed by evaluating the maximal cut. We will see
an explicit example of this later on in Section~\ref{sec:ellipticbox}.

Let us see how this works with a very simple example.
We consider the one-loop massive bubble
\begin{equation}
\Bub(d;s,m_1^2,m_2^2) =\adjustbox{valign=m}{\includegraphics[scale=0.8,trim=0 0 0.2cm -0.14cm]{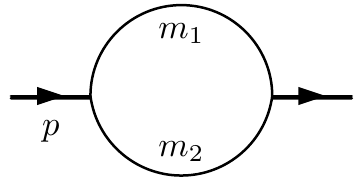} }= \int \D^d k \frac{1}{(k^2-m_1^2)((k-p)^2-m_2^2)}\,,
\end{equation}
where $p^2 = s.$
The latter satisfies very simple differential equations in $s$, $m_1^2$ and $m_2^2$.
The one in $s$, for example, reads
\begin{align} \frac{d}{ds} \Bub(d;s,m_1^2,m_2^2) &= 
   \frac{1}{2}
   \left( \frac{3-d}{(m_1-m_2)^2-s}+\frac{3-d}{(m_1+m_2)^2-s}+\frac{2-d}{s} \right)\Bub(d;s,m_1^2,m_2^2) \nonumber \\
  &+ G(d;s,m_1^2,m_2^2) \ , 
   \label{eq:EqBd} \end{align} 
where $G(d;s,m_1^2,m_2^2)$ is the inhomogeneous part which depends only on the tadpoles
\begin{align}
\Tad(d;m) = \int \D^d k \frac{1}{k^2-m^2}  = \frac{m^{d-2}}{(d-2)(d-4)}\,.\label{eq:tad}
\end{align}
Eq.\eqref{eq:tad} defines also our integration measure, whose exact value will nevertheless
be irrelevant in this context.
The homogeneous part of the equation~\eqref{eq:EqBd} reads (we use the suffix $_H$ to indicate
that we are neglecting all inhomogeneous terms)
\begin{align} \frac{d}{ds} \Bub_H(d;s,m_1^2,m_2^2) = 
   \frac{1}{2}
   \left(\frac{3-d}{(m_1-m_2)^2-s}+\frac{3-d}{(m_1+m_2)^2-s}+\frac{2-d}{s}\right)\Bub_H(d;s,m_1^2,m_2^2) 
   \label{eq:EqBdH} \end{align} 
and it admits the solution for generic $d$
\begin{align}
\Bub_H(d;s,m_1^2,m_2^2) = c_1\, (-s)^{1-\frac{d}{2}} \left(m_1^4-2 m_1^2
   \left(m_2^2+s\right)+\left(m_2^2-s\right)^2\right)^{\frac{d-3}{2}}\,, \label{eq:solBHd}
\end{align}
where $c_1$ is an irrelevant multiplicative constant.
By the considerations above,~Eq.\eqref{eq:solBHd} is also the $d$-dimensional 
maximal cut of the one-loop bubble.
Clearly, in this case, the maximal cut coincides with the cut in the $s$-channel, which reads
\begin{align}
\Cut\left(\adjustbox{valign=m}{\includegraphics[scale=0.8,trim=0 0 0.2cm -0.14cm]{fig/fig1Lbubble.pdf} }\right)=
\int \D^d k\;  \delta(k^2-m^2) \delta((k-p)^2-m^2)\,.
\end{align}
The cut is straightforward to compute in the frame $p^\mu = (\sqrt{s}, \vec{0})$
and one immediately obtains
\begin{align}
\Cut\left(\adjustbox{valign=m}{\includegraphics[scale=0.8,trim=0 0 0.2cm -0.14cm]{fig/fig1Lbubble.pdf} }\right)=
c_2\, (-s)^{1-\frac{d}{2}} \left(m_1^4-2 m_1^2
   \left(m_2^2+s\right)+\left(m_2^2-s\right)^2\right)^{\frac{d-3}{2}}\,, \label{eq:cutBd}
\end{align}
where $c_2$ is another constant whose exact value is irrelevant.
Comparing Eq.~\eqref{eq:solBHd} and Eq.~\eqref{eq:cutBd} it is clear that, up to an irrelevant 
multiplicative constant, the solution of the homogeneous equation coincides with the maximal
cut of the graph.\\
A comment is in order. If a graph depends on more than one scale, like in the case above,
the maximal cut provides the homogeneous
solution of all differential equations in all Mandelstam invariants.
In general, solving only the homogeneous equation in one of the invariants
cannot capture the full dependence on all the remaining scales. For example,
in the case of the one-loop bubble analyzed above,
if we had solved only its homogeneous differential equations in $\partial/\partial m_1^2$, 
we would have not been able to determine the overall
dependence on $(-s)^{(1-d/2)}$ in Eq.~\eqref{eq:cutBd}.
Therefore, in order to get the full answer for the maximal cut we must solve
all homogeneous differential equations in all Mandelstam invariants.
While this example is straightforward, it contains most of the features of the more 
complicated examples that we will study below. 

Having retained full dependence on $d$ in the example above might look overly complicated.
Indeed, for physical applications,
we will be eventually interested in the solution of the system of differential equations as 
a Laurent series in $\epsilon = (4-d)/2$.
The discussion above applies of course for any integer values of $d$ as well and, in
particular, for $d = 4$.
If we go back to the general equation~\eqref{eq:gendeq}, put $d = 4-2\epsilon$ and expand it
in $\epsilon$, we will quite in general end up with a system of equations in the form
\begin{align}
\partial_x \, m_i(\epsilon; x) = H_{ij}(x)\,m_{j}(\epsilon; x) 
+ \sum_{k=1}^\infty \, \epsilon^k\, H_{ij}^{(k)}(x)\, m_{j}(\epsilon; x) + G_i(\epsilon; x).
\label{eq:gendeqexp}
\end{align}
Since we are interested in solving~\eqref{eq:gendeqexp} as Laurent series in $\epsilon$,
the crucial step consists now in solving the homogeneous
system evaluated for $\epsilon = 0$
\begin{align}
\partial_x \, h_i( x) = H_{ij}(x)\,h_{j}(x) \,.
\label{eq:homdeqd4}
\end{align}
Once the homogeneous solution is known, one can write order by order in $\epsilon$
an integral representation for the inhomogeneous solution, which will
then be represented in terms of iterated integrals.
Naively, we expect that solving~\eqref{eq:homdeqd4} must be substantially simpler
than retaining full dependence on $d$. While this is indeed true, 
there exists no general algorithm to solve~\eqref{eq:homdeqd4} 
if more than one equation is coupled, even for $d=4$.
It is sometimes useful to rephrase~\eqref{eq:homdeqd4} as a $N$-th order differential equation
for any of the functions $h_j(x)$ and try to solve the latter.
As of today, a limited number of two-loop examples are known which require the solution of 
second order differential equations. In all these cases, a solution could be found
in terms of complete
elliptic integrals of the first and second kind with complicated arguments and irrational
prefactors. The best known example is the two-loop massive 
sunrise graph~\cite{Laporta:2004rb,Adams:2013nia,Remiddi:2013joa,Adams:2014vja,Adams:2015gva,Adams:2015ydq},
while more recently other, in some cases unrelated, examples have been 
worked out~\cite{Aglietti:2007as,CaronHuot:2012ab,Remiddi:2016gno,Adams:2016xah,Bonciani:2016qxi}.
In all these cases, once the homogeneous solution was determined it became possible
to write useful integral representations for the complete result.

Until now these results have been obtained with a case by case analysis.
In this respect, very recently an interesting approach to solve a second order differential 
equation in terms of elliptic integrals has been proposed in~\cite{Bonciani:2016qxi}.
The latter is based on the possibility of reparametrizing the differental equations in terms of 
a suitably chosen parameter, in terms of which its solution in the form of elliptic
integrals becomes manifest.

We will follow here a complementary approach. 
We will show that, also in complicated cases which require the introduction of elliptic
integrals, the solution of the homogeneous equation for $\epsilon=0$
can be rather simply obtained from the maximal cut of the integral evaluated in $d=4$.
We will work out explicitly different examples, including the double box considered
in~\cite{Bonciani:2016qxi}.
While this will allow us to produce equivalent results, we believe that our approach has
an interest on its own, in particular since it can be at least in principle extended to any other
arbitrarily complicated example. 
As we will see, the main limitation of our approach is the possibility of performing 
explicitly the integrals
over the residual component of the loop momenta, after all $\delta$-functions stemming
from the propagators have been integrated out.
This can potentially be a serious limitation. Nevertheless,
we will show that in many cases this is not a problem and the calculation
can be organized so to be left with only one residual one-dimensional integration.
The complexity of this last integration depends, of course, on the integral under consideration.
In simple cases which can be solved in terms of multiple polylogarithms, it can be usually
performed in terms of simple rational functions or square roots of
rational functions. Whenever the result involves
elliptic function, instead,  one can use a quite general change of variable 
in order to rephrase
the result in standard form in terms of complete elliptic integrals.
Finally, even if the last integrations cannot be performed analytically,
this method allows to obtain relatively compact integral representations of one
homogeneous solution of a complicated higher order differential equation,
which is a very useful piece of information towards its complete solution.

In what follows we will see how the ideas described here work in practice in different examples.
We will start analyzing more in detail a one-loop example, and move then to
more interesting two-loop cases.

\section{A one-loop three-point function} 
\label{sec:oneloop} \setcounter{equation}{0} 
\numberwithin{equation}{section} 
At one loop, for every topology, one finds always at most one master integral, which
implies that one-loop integrals satisfy at most first order linear differential equations.
It is then usually straightforward to integrate the homogeneous part of the 
differential equation by quadrature. According to the discussion above, this
has to turn out to be identical to the computation of the maximal cut
for the integral under consideration.
Despite the simplicity of the calculations, we will do this here explicitly 
for an interesting example. 
This will allow us, on the one hand, to show 
some features of the procedure and, on the other, to obtain some useful 
formulas that we will recycle for the more interesting two-loop examples discussed below.

Let us consider a one-loop triangle with two massive internal propagators
and three off-shell external legs
\begin{align}
\Tri(d;q^2,q_1^2,q_2^2,m_a^2,m_b^2) &= \adjustbox{valign=m}{\includegraphics[scale=0.8,trim=0 0 0 -0.27cm]{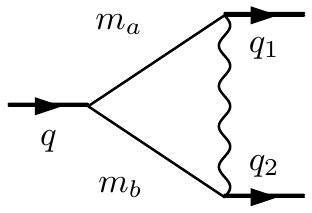} }
\nonumber \\
&=
\int \D^d k \frac{1}{(k^2-m_a^2)(k-q_1)^2((k-q_1-q_2)^2-m_b^2)}\,,
\end{align}
where in total generality $q^2 \neq q_1^2 \neq q_2^2\neq 0$.
In this case it is very simple to compute the maximal cut of the graph,
in particular in $d=4$:
\\
\begin{align}
\Cut\left(\adjustbox{valign=m}{\includegraphics[scale=0.7,trim=0 0 0 -0.27cm]{fig/fig1LTrima.pdf} }\right)&=
\int \D^4 k \delta(k^2-m_a^2) \delta((k-q_1)^2) \delta((k-q_1-q_2)^2-m_b^2)\,.
\end{align}
We go to the reference frame $q^\mu = (\sqrt{q^2}, \vec{0})$, 
$q_1^\mu = (E_1,0,0,q_{1z})$, $q_2^\mu = (E_2,0,0,-q_{1z})$ with
$$q_{1z} = \frac{\sqrt{(q^2+q_1^2-q_2^2)^2 - 4\ q^2\, q_1^2}}{2 \sqrt{q^2}}$$
and 
obtain immediately
(we neglect at every step irrelevant overall numerical constants)
\begin{align}
\Cut &\left(\adjustbox{valign=m}{\includegraphics[scale=0.7,trim=0 0 0 -0.27cm]{fig/fig1LTrima.pdf} }\right)=
\int_{-1}^{1}\!dz \int_0^\infty\!  d  \kk \kk^2
\int_{-\infty}^{+\infty} \! d k_0 
\delta(k^2-m_a^2) \delta((k-q_1)^2) \delta((k-q_1-q_2)^2-m_b^2)\, \nonumber \\
&=
\int_{-1}^{1} dz\, \int_0^\infty  d  \kk\, \frac{\kk^2}{\omega_a}\;
\delta(q_1^2 + m_a^2 -2 \omega_a\, E_1 + 2 \kk \, q_{1z}\, z)\, \delta(q^2 + m_a^2 - m_b^2 - 2\, \omega_a \sqrt{q^2})\,,
\end{align}
where we defined $\kk = |\vec{k}|$ and $\omega_a = \sqrt{\kk^2 + m_a^2}$\,.
Performing the integral in $d\kk = \omega_a/\kk\, d\omega_a$ we get
\begin{align}
\Cut &\left(\adjustbox{valign=m}{\includegraphics[scale=0.7,trim=0 0 0 -0.27cm]{fig/fig1LTrima.pdf} }\right)
 = \frac{1}{q_{1z}\,\sqrt{q^2}}
\int_{-1}^{1} dz\, \;
\delta\left( \frac{q_1^2 + m_a^2 -2 \bar{\omega}_a\, E_1}{2  \,q_{1z}\, \sqrt{\bar{\omega}_a^2-m_a^2} } 
+  z \right)\, ,
\label{eq:rescutTri1loop1}
\end{align}
where we have fixed $\bar{\omega}_a = (q^2 + m_a^2-m_b^2)/2/\sqrt{q^2}\,.$
Given that there exists a real or complex value for the momenta where the
$\delta$-function in~\eqref{eq:rescutTri1loop1} has support\footnote{As stated previously, 
it is a well-known fact
that, in general, a solution for the maximal cut exists only if we allow the external
invariants to assume complex values.}, we are left with
\begin{align}
\Cut &\left(\adjustbox{valign=m}{\includegraphics[scale=0.7,trim=0 0 0 -0.27cm]{fig/fig1LTrima.pdf} }\right)
=
\frac{1}{\sqrt{(q^2+q_1^2-q_2^2)^2 - 4\ q^2\, q_1^2}}\,.
 \label{eq:rescutTri1loop}
\end{align}

Now, following our argument, this cut must be proportional to the solution of the
homogeneous differential equations satisfied by $\Tri(4;q^2,q_1^2,q_2^2,m_a^2,m_b^2)$.
It is very simple to derive the differential equations satisfied by the latter in all external invariants.
As exemplification we write down here the  homogeneous differential equation in $q^2$ 
for generic $d$, which reads (again, we use the suffix $_H$ to indicate that we are neglecting all
inhomogeneous terms in the differential equation)
\begin{align}
\frac{\partial}{\partial q^2} \Tri_H(d;q^2,q_1^2,q_2^2,m_a^2,m_b^2) &= \frac{(q_1^2+q_2^2-q^2)}{q_1^4-2 q_1^2(q_2^2+q^2)+(q_2^2-q^2)^2}  \Tri_H(d;q^2,q_1^2,q_2^2,m_a^2,m_b^2) 
\nonumber \\
&+  (d-4)\, C(q^2,q_1^2,q_2^2,m_a^2,m_b^2)\, \Tri_H(d;q^2,q_1^2,q_2^2,m_a^2,m_b^2)\,, 
\label{eq:deqTri1loop}
\end{align}
where $C(q^2,q_1^2,q_2^2,m_a^2,m_b^2)$ is a cumbersome rational function.
In the simpler case where $m_b = m_a$ the latter reads
\begin{align}
C(q^2,q_1^2,q_2^2,m_a^2) &= C(q^2,q_1^2,q_2^2,m_a^2,m_a^2) \nonumber \\
&= \frac{ ((m_a^2+q_1^2) (q_1^2-q_2^2)+s (m_a^2-q_1^2)) ((m_a^2+q_2^2) (q_1^2-q_2^2)+s
   (q_2^2-m_a^2))}{2 \left(q_1^4-2 q_1^2 (q_2^2+s)+(q_2^2-s)^2\right) \left(s (m_a^2-q_1^2)
   (m_a^2-q_2^2)+m_a^2 (q_1^2-q_2^2)^2\right)}\,.
\end{align}
Putting $d=4$ in Eq.~\eqref{eq:deqTri1loop}, one sees that the entire dependence on the
masses $m_a$ and $m_b$ disappears and the solution reads
\begin{align}
\Tri_H(4;q^2,q_1^2,q_2^2,m_a^2,m_b^2)  =
\frac{c}{\sqrt{(q^2+q_1^2-q_2^2)^2 - 4\ q^2\, q_1^2}}\,,
 \label{eq:rescutTri1loop2}
\end{align}
where $c$ is an integration constant. 
This is of course in perfect agreement with the result of the maximal cut calculation, Eq.~\eqref{eq:deqTri1loop}. 

Two comments are in order here. First of all, we could have of course solved Eq.~\eqref{eq:deqTri1loop}
retaining full dependence in $d$. The result is relatively compact and reads
\begin{align}
& \Tri_H(d;q^2,q_1^2,q_2^2,m_a^2,m_b^2)  = 
\left(q_1^4-2 q_1^2 (q_2^2+s)+(q_2^2-s)^2\right)^{\frac{3-d}{2}} \nonumber \\
&\times\left(m_a^2 (m_b^2 (q_1^2+q_2^2-s)+q_2^2 (q_1^2-q_2^2+s))+q_1^2 s
   (m_b^2-q_2^2)-m_b^2 q_1^2 (m_b^2+q_1^2-q_2^2) - m_a^4\,q_2^2\right)^{\frac{d-4}{2}}
 \label{eq:rescutTri1loopd}
\end{align}
Again, up to an overall constant, this is also the result of the calculation of the 
$d$-dimensional maximal cut 
of this triangle. 

A second interesting feature of our result is that, in the limit $d \to 4$, 
it does not depend either on $m_a$ 
or $m_b$. This has some important consequences. 
Let us consider a similar integral, but with one of the two massive propagators squared,
for example
\begin{align}
\Tri_{2a}(d;q^2,q_1^2,q_2^2,m_a^2,m_b^2) &=  \adjustbox{valign=m}{\includegraphics[scale=0.8,trim=0 0 0 -0.27cm]{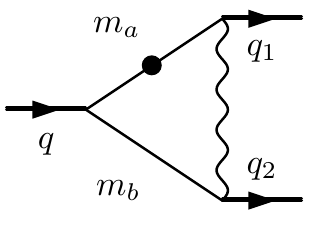} }
 \nonumber \\
&=
\int \D^d k \frac{1}{(k^2-m_a^2)^2(k-q_1)^2((k-q_1-q_2)^2-m_b^2)}\,,
\end{align}
such that
$$\Tri_{2a}(d;q^2,q_1^2,q_2^2,m_a^2,m_b^2) = \frac{\partial}{\partial\, m_a^2}
\Tri(d;q^2,q_1^2,q_2^2,m_a^2,m_b^2)\,.$$
Following our prescriptions for computing the maximal cut of this integral, 
we define simply
\begin{align}
\Cut \left( \adjustbox{valign=m}{\includegraphics[scale=0.8,trim=0 0 0 -0.27cm]{fig/fig1LTrimadot.pdf} }
 \right)
= \frac{\partial}{\partial\, m_a^2} \Cut \left( \adjustbox{valign=m}{\includegraphics[scale=0.8,trim=0 0 0 -0.27cm]{fig/fig1LTrima.pdf} }
\right)\,,
\end{align}
so that for $d=4$ we obtain immediately
\begin{align}
\Cut \left(\adjustbox{valign=m}{\includegraphics[scale=0.8,trim=0 0 0 -0.27cm]{fig/fig1LTrimadot.pdf} } 
 \right)
\propto \frac{\partial}{\partial\, m_a^2} \frac{1}{\sqrt{(q^2+q_1^2-q_2^2)^2 - 4\ q^2\, q_1^2}} = 0\,.
\label{eq:rescutTri2a}
\end{align}
The very same thing is true if we put a dot on the other massive propagator.
We should stress here once more that this result is valid only for $d=4$ identically.
It is clear from Eq.~\eqref{eq:rescutTri1loopd} that taking a derivative w.r.t.
$m_a^2$ or $m_b^2$ would produce an overall factor $(d-4)$ which goes to 
zero as $d \to 4$.

A similar conclusion can be drawn inspecting the integration by parts identities.
For ease of writing, we consider again the case $m_b=m_a$.
Using IBPs one can show that
\begin{align}
 \adjustbox{valign=m}{\includegraphics[scale=0.8,trim=0 0 0 -0.27cm]{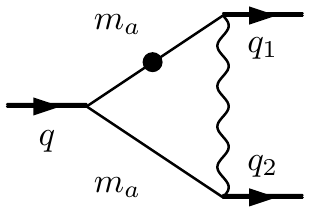} }&=
\frac{(d-4) ((m_a^2+q_2^2) (q_2^2-q_1^2)+q^2 (m_a^2-q_2^2))}{2 q^2 (m_a^2-q_1^2) (m_a^2-q_2^2)+2 m_a^2
   (q_1^2-q_2^2)^2}\;\;
  \adjustbox{valign=m}{\includegraphics[scale=0.8,trim=0 0 0 -0.27cm]{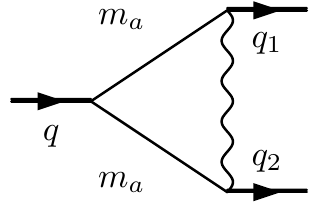} }
 \nonumber \\&\nonumber \\
 &+ {\rm subtopologies}\,. \label{eq:TriIBP}
\end{align}
Of course a similar but more cumbersome expression can be derived for $m_b \neq m_a$.
Clearly, computing the maximal cut on the right- and left-hand-side of Eq.~\eqref{eq:TriIBP}
all subtopologies drop and we are left with
\begin{align}
&\Cut\left(\adjustbox{valign=m}{\includegraphics[scale=0.8,trim=0 0 0 -0.27cm]{fig/fig1LTrimaadot.pdf} }
 \right) = 
 \frac{(d-4) ((m_a^2+q_2^2) (q_2^2-q_1^2)+q^2 (m_a^2-q_2^2))}{2 q^2 (m_a^2-q_1^2) (m_a^2-q_2^2)+2 m_a^2
   (q_1^2-q_2^2)^2}\;\;
 \Cut \left( \adjustbox{valign=m}{\includegraphics[scale=0.8,trim=0 0 0 -0.27cm]{fig/fig1LTrimaa.pdf} }
  \right) \,,
 \label{eq:TriIBPcut}
\end{align}
which, given Eq.~\eqref{eq:rescutTri1loopd}, allows to compute the $d$-dimensional
cut of the triangle with a squared propagator.
Putting then $d=4$ one is left again with
\begin{align}
&\Cut\left( \adjustbox{valign=m}{\includegraphics[scale=0.8,trim=0 0 0 -0.27cm]{fig/fig1LTrimadot.pdf} }
 \right) \Bigg|_{d=4} = 0\,,\end{align}
consistently with~\eqref{eq:rescutTri2a}.

\section{Two-loop examples}
\label{sec:twoloop} \setcounter{equation}{0} 
\numberwithin{equation}{section} 
At two-loop order the situation is, indeed, much more interesting.
In this case, one often has to consider topologies of Feynman integrals
which are reduced to two or more master integrals. The latter will therefore
satisfy systems of coupled differential equations, whose solution becomes
much less straightforward.
In this case, the first step consists in finding a solution of the homogeneous 
system. Once the latter is known, one can proceed using Euler's variation
of the constants to write down an integral representation for the full
solution.
In what follows we will considered two examples of two-loop Feynman graphs 
which fulfill second order differential equations and 
 show explicitly how the maximal cut allows us to
find the required homogeneous solutions.

\subsection{A non-planar two-loop triangle}
Let us start considering a two-loop
non-planar triangle with two off-shell legs and four massive 
propagators which enters the non-planar
corrections to H+j production through a massive fermion loop.
We define our integral family as follows
\begin{align}
&  \adjustbox{valign=m}{\includegraphics[scale=0.7,trim=0 0 0 -0.27cm]{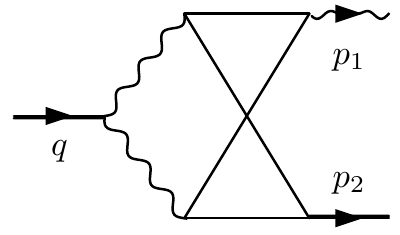} }
= I_{a_1,a_2,a_3,a_4,a_5,a_6,a_7} \Big|_{a_7 < 0} \nonumber \\
& = \int \, 
\frac{ \D^dk\, \D^d l \;\; (k^2)^{-a_7}}{[(k-p_1)^2]^{a_1} [(l-p_1)^2-m^2]^{a_2} [(k+p_2)^2]^{a_3}
[(k-l+p_2)-m^2]^{a_4} [(k-l)-m^2]^{a_5}[l^2-m^2]^{a_6}}\,,   \label{eq:topo}
\end{align}
with  $p_1^2 = 0$, $p_2^2 \neq 0$ and $q^2 = (p_1+p_2)^2 = s  \neq 0 $\,.
The integrals~\eqref{eq:topo} can be reduced to two master integrals that we choose as
\begin{align}
\I_{1} &= (s-p_2^2)^2 I_{1, 1, 1, 1, 1, 1, 0}\,, \qquad 
\I_{2} = \frac{(s-p_2^2)^2 (s + 16 m^2)}{2 (1 + 2 \e)} I_{1, 2, 1, 1, 1, 1, 0}\,.
\label{eq:basis2}
\end{align}
Neglecting all subtopologies, the latter satisfy the following two sets of coupled differential equations in the variables $x=-s/m^2$ and $y=-p_2^2/m^2$,
\begin{align}
\frac{d}{dx} \left(\begin{matrix} \I_{1} \\ \I_{2} \end{matrix} \right) =&
B_x(x,y) \left(\begin{matrix} \I_{1} \\ \I_{2} \end{matrix} \right) 
+ \epsilon\, D_x(x,y) \left(\begin{matrix} \I_{1} \\ \I_{2} \end{matrix} \right),\nn
\frac{d}{dy} \left(\begin{matrix} \I_{1} \\ \I_{2} \end{matrix} \right) =&
B_y(x,y) \left(\begin{matrix} \I_{1} \\ \I_{2} \end{matrix} \right) 
+ \epsilon\, D_y(x,y) \left(\begin{matrix} \I_{1} \\ \I_{2} \end{matrix} \right)
 \label{eq:topsys2}
\end{align}
where $B_{x,y}(x,y)$ and $D_{x,y}(x,y)$ are $2 \times 2$ matrices that do not depend on $\epsilon$, given by
\begin{align}
B_x(x,y) =& \left( \begin{array}{ccc} 0 & & \frac{16 x}{(x-16) (y-x)^2}+\frac{8}{(x-16) (y-x)} \\&\\
                                        \frac{(x-16) (x+y)}{2 x \left(x^2-2 x y-16 x+y^2\right)} & &-\frac{2 (x-y-8)}{x^2-2 x y-16 x+y^2}-\frac{3}{y-x}+\frac{16}{(x-16) x}  \end{array}\right)\,, 
                                   \nn
                                   \nn
D_x(x,y) = &\left( \begin{array}{ccc} \frac{2}{y-x} & & \frac{32 x}{(x-16) (y-x)^2}+\frac{16}{(x-16) (y-x)} \\&\\
                                       \frac{2 (x-16) (x+y)}{x \left(x^2-2 x y-16 x+y^2\right)} & & -\frac{2 (x-y-8)}{x^2-2 x y-16 x+y^2}-\frac{4}{y-x}-\frac{2}{x} \end{array}\right)\,,
                                   \nn
                                   \nn
B_y(x,y) =& \left( \begin{array}{ccc} 0 & &-\frac{16 x}{(x-16) (y-x)^2} \\&\\
                                        \frac{16-x}{x^2-2 x y-16 x+y^2} & &\frac{2 (x-y)}{x^2-2 x y-16 x+y^2}+\frac{3}{y-x}  \end{array}\right)\,, 
                                   \nn
                                   \nn
D_y(x,y) = &\left( \begin{array}{ccc} \frac{2}{x-y} & &-\frac{32 x}{(x-16) (x-y)^2} \\&\\
                                       -\frac{4 (x-16)}{x^2-2 x (y+8)+y^2} & & -\frac{2 x^2-4 x y-64 x+2 y^2}{(x-y) \left(x^2-2 x (y+8)+y^2\right)} \end{array}\right)\,.                                       
\end{align}
In $d=4$, ($\epsilon = 0)$ the systems become
\begin{align}
\frac{d}{dx} \left(\begin{matrix} \I_{1} \\ \I_{2} \end{matrix} \right) =&
B_x(x,y) \left(\begin{matrix} \I_{1} \\ \I_{2} \end{matrix} \right)\,,\qquad
\frac{d}{dy} \left(\begin{matrix} \I_{1} \\ \I_{2} \end{matrix} \right) =
B_y(x,y) \left(\begin{matrix} \I_{1} \\ \I_{2} \end{matrix} \right)\,
\label{eq:topsys42}
\end{align}
and they can be rephrased as second order differential equations
for one of the two master integrals. For $\I_1$ they read
\begin{align}
\frac{d^2 \, \I_1(x,y) }{dx^2} &+\left(\frac{1}{y-x}-\frac{1}{x+y}+\frac{1}{x}+\frac{2 (x-y-8)}{x^2-2 x y-16 x+y^2}\right)\frac{d \, \I_1(x,y) }{dx}\nn 
&+\left(\frac{1}{x (y-x)}+\frac{1}{(y-x)^2}-\frac{y+4}{x \left(x^2-2 x y-16 x+y^2\right)}\right)\I_1(x,y) = 0\,
\label{eq:secdeqm2x}
\end{align}
and
\begin{align}
\frac{d^2 \, \I_1(x,y) }{dy^2} &-\left(\frac{1}{y-x}+\frac{2 (x-y)}{x^2-2 x y-16 x+y^2}\right)\frac{d \, \I_1(x,y) }{dy}\nn 
&+\left(\frac{1}{(y-x)^2}-\frac{1}{x^2-2 x y-16 x+y^2}\right)\I_1(x,y) = 0\,.
\label{eq:secdeqm2y}
\end{align}
Instead of trying to solve these equations directly, 
we compute the maximal cut of $\I_1$ in order to determine a first homogeneous
solution. 
$\I_1$ can be written as
\begin{align}
\I_1 &= \int \frac{ \D^dk}{(k-p_1)^2\, (k+p_2)^2\,} 
\int \frac{\D^d l}{ (l^2-m^2)  [(k-l)-m^2] [(l-p_1)^2-m^2]\, [(k-l+p_2)-m^2]\,}\,.
\label{eq:tocut2}
\end{align}
The integral in $\D^dl$ is a one-loop box with four massive propagators and three
off-shell external legs
\begin{equation}
 \adjustbox{valign=m}{\includegraphics[scale=0.8,trim=0 0 0 -0.1cm]{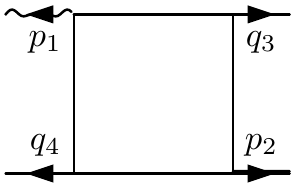} } 
=
 \int \frac{\D^d l}{ (l^2-m^2)  [(k-l)^2-m^2] [(l-p_1)^2-m^2]\, [(k-l+p_2)^2-m^2]\,} \,,
 \label{eq:onebox}
\end{equation}
where $p_1$ and $p_2$ are defined above, while $q_3 = k-p_1$ and $q_4= -k-p_2$.

In order to compute the maximal cut of $\I_1$, it is convenient to start by first cutting 
the one-loop box Eq.~\eqref{eq:onebox}.
Since we have already spelled out the relation between maximal cut and homogeneous
differential equation, we can  avoid to compute this cut by direct integration over the
$\delta$-functions.
Instead, we derive the homogeneous differential equations satisfied by the
latter in all external invariants and solve them
for $d=4$.
The result provides us with the maximal cut, up to an overall irrelevant constant.
We obtain
\begin{align}
&\Cut \left( \adjustbox{valign=m}{\includegraphics[scale=0.8,trim=0 0 0.1cm -0.1cm]{fig/fig1LBox.pdf} }
 \right) = \nonumber \\
&= \frac{c}{\sqrt{ (q_3^2 q_4^2-t u)^2 + 
4 m^2 \left(p_2^2 (q_3^2-t) (q_4^2-u)-q_3^4 q_4^2
  -q_3^2 \left(q_4^4-q_4^2 (t+u)-t u\right)+t u (q_4^2-t-u)\right)}}
\,,   \label{eq:boxcut2} 
\end{align}
where we have defined $t=(p_1-q_3)^2$ and  $u=(p_2-q_3)^2$. Substituting this into~\eqref{eq:tocut2} and localizing the contour for the
remaining two propagators we are left with
\begin{align}
\Cut\left(\adjustbox{valign=m}{\includegraphics[scale=0.7,trim=0 0 0 -0.25cm]{fig/fig2LTriNP.pdf} }\right)
& = (s-p_2^2)^2  \int \, 
\D^dk \frac{\delta(k^2) \delta((k-p_1-p_2)^2) }{\sqrt{t u (t u - 4 m^2 (t+u - p_2^2))}} \,,
\end{align}
where we used the condition $q_3^2= q_4^2 =0$.
In order to compute the second integral, we go to the center of mass frame and perform the integration in $k_0$ and $\bar{k}$ 
and obtain
\begin{align}
\Cut\left(\adjustbox{valign=m}{\includegraphics[scale=0.7,trim=0 0 0 -0.25cm]{fig/fig2LTriNP.pdf} }\right)
&=
 (s-p_2^2)\int_{-1}^{1}dz\, \frac{1}{\sqrt{ \left(1-z^2\right) 
 \left( 16 m^2 s + \left(1-z^2\right) (s-p_2^2)^2\right)}} \nonumber \\
 &= \frac{(s-p_2^2)}{ \sqrt{p_2^4-2\,p_2^2\,s + s (s+16\, m^2)}}
 \int_{-1}^{1}dz\, \frac{1}{\sqrt{ \left(1-z^2\right) 
 \left( 1 - w\, z^2\right)}}\nonumber \\
 &= \frac{(s-p_2^2)}{\sqrt{p_2^4-2\,p_2^2\,s + s (s+16\, m^2)}}
 \EK(w),
 \label{eq:cut2}
\end{align}
where we used the definition of the complete elliptic integral of the first kind
\begin{equation}
\EK(x) = \int_0^1 \frac{dt}{\sqrt{(1-t^2)(1-\,x\,t^2)}} \,, \quad 
\mbox{for} \quad x \in \mathbb{C} \quad \mbox{and}\quad \Re{(x)} < 1\,, 
\end{equation}
and we defined
$$w = \frac{(s-p_2^2)^2}{p_2^4-2\,p_2^2\,s + s (s+16\, m^2) }\,.$$
In terms of the adimensional variables $x$ and $y$, the maximal cut becomes 
(neglecting once more overall constant factors)
\begin{align}
\Cut(\I_1) &\propto F_1(x,y) \equiv
\frac{(x-y) }{\sqrt{x^2-2 x (y+8)+y^2}}\EK \left(\frac{(x-y)^2}{x^2-2 (y+8) x+y^2}\right).
 \label{eq:solcut2}
\end{align}
By direct computation one can check that $F_1(x,y)$ solves both second order differential equations \eqref{eq:secdeqm2x} and \eqref{eq:secdeqm2y}. Finally, the properties of elliptic functions allow us to write a second independent solutions as
\begin{align}
F_2(x,y) &\equiv
\frac{(x-y)}{\sqrt{x^2-2 x (y+8)+y^2}}\EK\left(1-w \right) \nonumber \\
&=\frac{(x-y)}{\sqrt{x^2-2 x (y+8)+y^2}}\EK\left(-\frac{16 x}{x^2-2 (y+8) x+y^2}\right).
\end{align}
We observe that in the limit $p_2^2 \to 0$ we have
\begin{align}
\Cut(\I_1, p_2^2\to 0) &\propto
 \frac{s}{ \sqrt{s (s+16\, m^2)}}
 \EK\left(\frac{s}{s+16\, m^2}  \right),
\end{align}
which provides the homogeneous solution for the differential equations
of the corresponding non-planar two-loop triangle with only one off-shell leg.
This solution can be then analytically continued to all physically
relevant phase-space region and can be used to write down a one-fold integral
representation for the two-loop triangle. 
This problem will be considered elsewhere~\cite{TwoLoopNPL}.

%%%%%%%%%%%%%%%%%%%
%%%%%%%%%%%%%%%%%%
\subsection{An elliptic planar double box}
\label{sec:ellipticbox}
As a last non-trivial example, we consider the planar double-box
computed in~\cite{Bonciani:2016qxi}, which corresponds to the integral family
\begin{align}
\adjustbox{valign=m}{\includegraphics[scale=0.8,trim=0 0 0.1cm -0.1cm]{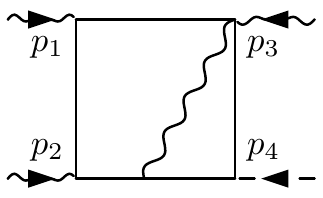} }=&\, I_{a_1\,a_2\,a_3\,a_4\,a_5\,a_6\,a_7\,a_8\,a_9}\Big|_{a_{7,8,9} < 0} 
=\quad \int \D^d\, k \D^d \,l  \frac{(k\cdot p_3)^{-a_7}(l\cdot p_1)^{-a_8}(l\cdot p_2)^{-a_9}}{D_1^{a_1}D_2^{a_2}\,D_3^{a_3}\,D_4^{a_4}\,D_5^{a_5}\,D_6^{a_6}},
\label{eq:intfambox}
\end{align}
and the propagators are defined as
\begin{align}
D_1=&k^2-m^2,\qquad\qquad\qquad \quad
D_2=(k-p_1)^2-m^2,\nn
D_3=&(k-p_1-p_2)^2-m^2,\qquad
D_4=(k-l+p_3)^2,\nn
D_5=&l^2-m^2,\qquad\qquad\qquad\quad\,
D_6=(l-p_1-p_2-p_3)^2-m^2.
\end{align}
We stress that, since we are interested in the six-denominator topology only, we restrict the indices $a_{7,8,9}$ to negative values. The external kinematics is chosen to be $p_1^2=p_2^2=p_3^2=0$, $p_4^2=(p_1+p_2+p_3)^2=m_h^2$ and the Mandelstam invariants are given by
\begin{align}
s=(p_1+p_2)^2\,,\quad t=(p_1+p_3)^2\,,\quad u=(p_2+p_3)^2=m_h^2-s-t.
\end{align}
The IBPs reduction of the integral family \eqref{eq:intfambox} returns four master integrals, which we choose to be
\begin{align}
\I_1 =& I_{1\,1\,1\,1\,1\,1\,0\,0\,0}\,,\qquad \I_2=I_{1\,2\,1\,1\,1\,1\,0\,0\,0}\nn
\I_3=& I_{1\,1\,1\,1\,2\,1\,0\,0\,0}\,,\qquad \I_4=I_{1\,1\,1\,1\,1\,1\,-1\,0\,0}.
\end{align}
The master integrals $\I_{i}$ fulfill systems of first order differential equations in the kinematic invariants. 
In addition, it can be verified that in $d=4$ the differential equations for $\I_4$ are completely decoupled. Therefore, in the four-dimensional limit we can restrict ourselves to study the homogeneous  systems for the first three master integrals, which read
\begin{align}
\begin{cases}
 \dfrac{\partial\,}{\partial x}\I_1(\bf{x})=&a_{11}(\textbf{x})\,\I_1(\textbf{x})+a_{12}(\textbf{x})\,\I_2(\textbf{x})+a_{13}(\textbf{x})\,\I_3(\textbf{x})\\
 \\
 \dfrac{\partial\,}{\partial x}\I_2(\textbf{x})=&a_{21}(\textbf{x})\,\I_1(\textbf{x})+a_{22}(\textbf{x})\,\I_2(\textbf{x})+a_{23}(\textbf{x})\,\I_3(\textbf{x})\\
 \\
 \dfrac{\partial\,}{\partial x}\I_3(\textbf{x})=&a_{33}(\textbf{x})\,\I_3(\textbf{x})\,,
 \end{cases}
 \label{eq:sysbox3}
 \end{align}
 where $x\in \textbf{x}=\{s,t,m_h^2,m^2\}$ and $a_{ij}(\textbf{x})$ are rational functions
 in the Mandelstam invariants. 
Following our argument, we expect that computing the maximal cut of the three
master integrals we will get, by construction, one set of homogeneous solutions
of the system~\eqref{eq:sysbox3}. 

Interestingly enough, it is easy to show that cutting all propagators produces a further simplification of the system. In fact, we observe that $\I_3$ contains as sub-loop the one-loop 
triangle with one dotted massive propagator studied in Section~\ref{sec:oneloop}.
There we saw that the latter has vanishing maximal cut in $d=4$. 
Hence, as a direct consequence of Eq.\eqref{eq:rescutTri2a}, we immediately get
 \begin{align}
 \Cut\left(\adjustbox{valign=m}{\includegraphics[scale=0.8,trim=0 0 0.1cm -0.1cm]{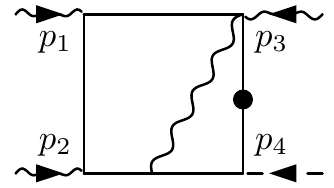} }\right)=&
 \int \D^4 k\, \delta(k^2-m^2) \delta((k-p_1)^2-m^2)\delta((k-p_1-p_2)^2-m^2)\nn\
&\times\,\Cut \left(  \adjustbox{valign=m}{\includegraphics[scale=0.7,trim=0 0 0 -0.17cm]{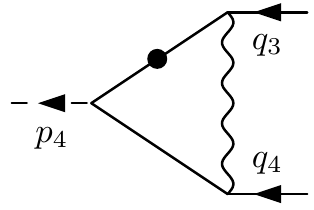} }
\right)=0.
\label{eq:cutboxI3}
 \end{align}
The vanishing of $\Cut\left(\I_3\right)$ implies that, when the systems \eqref{eq:sysbox3} are evaluated on the maximal cut, the last equation drops out and we are left with two coupled homogenous first order differential equations for $\I_1$ and $\I_2$,
\begin{align}
\begin{cases}
 \dfrac{\partial\,}{\partial x}\Cut\left(\I_1(\textbf{x})\right)=&a_{11}(\textbf{x})\,\Cut\left(\I_1(\textbf{x})\right)+a_{12}(\textbf{x})\,\Cut\left(\I_2(\textbf{x})\right)
 \\
 \\
 \dfrac{\partial\,}{\partial x}\Cut\left(\I_2(\textbf{x})\right)=&a_{21}(\textbf{x})\,\Cut\left(\I_1(\textbf{x})\right)+a_{22}(\textbf{x})\,\Cut\left(\I_2(\textbf{x})\right)\,.
 \\
 \end{cases}
  \label{eq:sysbox2}
 \end{align}
A comment on the relation between the solutions of the homogenous systems \eqref{eq:sysbox3} and the existence of a vanishing maximal cut is in order.
Since the differential equation for $\I_3$ is decoupled from $\I_1$ and $\I_2$, a zero maximal cut provides an obvious solution to it. However, $\Cut\left(\I_3\right)=0$ does not prevent the differential equation to admit other non-trivial solutions, which are not captured by the maximal cut. Nevertheless, this is not a problem since the differential equation
is decoupled and can be solved independently by quadrature.
We would like to stress here that the decoupling of the differential equation for a master integral in four dimensions and the vanishing of its maximal cut are indeed closely related. In fact, precisely a vanishing maximal cut could be seen as the hint of the decoupling of the differential equation for the corresponding master integral, since only a decoupled equation would be automatically satisfied by a zero solution without imposing strong constraints on the maximal cuts of the other two master integrals. \\
 
The systems of equations \eqref{eq:sysbox2} can then be rephrased as a second order differential equations for the maximal cut of one of two master integrals. For instance, if we choose $\I_1$ and we differentiate with respect to the internal mass, we have
 \begin{align}
 &\dfrac{\partial^2\,}{(\partial m^2)^2}\Cut \left(\I_1\right)=\nn
 &\frac{s^2 \left(48 m^4-16 m^2 (t+m_h^2)+(t-m_h^2)^2\right)+48 m^4 t^2+16 m^2 s t (6 m^2-t+m_h^2)}{m^2 \left((m_h^2-4
   m^2)^2 s^2+t^2 (s-4 m^2)^2-2 (4 m^2+m_h^2) s t (s-4 m^2)\right)}  \dfrac{\partial\,}{\partial m^2}\Cut\left(\I_1\right)\nn
   \nn
   &-\frac{2 \left(s^2 (-6 m^2+t+m_h^2)+s t (-12 m^2+t-m_h^2)-6 m^2 t^2\right)}{m^2 \left((m_h^2-4 m^2)^2
   s^2+t^2 (s-4 m^2)^2-2 (4 m^2+m_h^2) s t (s-4 m^2)\right)}\Cut\left(\I_1\right)
   \label{eq:2ndordm2}
 \end{align}
 and other three similar equations can be determined by taking derivatives with respect to $s$, $t$ and $m_h^2$. \\

 One of the two independent solutions of this set of second order differential equations can be found by direct computation of the maximal cut of $\I_1$. 
As in the previous example, we start by computing the maximal cut of one of the two sub-loops from the solution of the corresponding differential equations in $d=4$ and then integrate over the second loop momentum by solving explicitly the constraints imposed by the remaining $\delta$-functions. In this respect we observe that, in principle, one is completely free to choose the order in which the loop momenta are localized. Nevertheless, a wise choice can substantially
simplify the remaining integrals.
In this particular case, the two sub-loops correspond to a box with two adjacent off-shell legs and a triangle with three off-shell legs. If we started by cutting the box, we would be left with an integral over the four variables parametrizing the second loop momentum and only two $\delta$-functions constraining them. Therefore, in order to obtain a one-fold integral representation of the solution, we would need to perform an additional non-trivial integration.
On the other hand, if we localize the triangle integral first, the action of the three remaining $\delta$-functions would directly provide an expression of the maximal cut in terms of a one-dimensional integral.\\

We start then from the cut of the internal one-loop triangle which, using the results of Section~\ref{sec:oneloop}, reads
\begin{align}
&\Cut \left(\adjustbox{valign=m}{\includegraphics[scale=0.7,trim=0 0 0 -0.17cm]{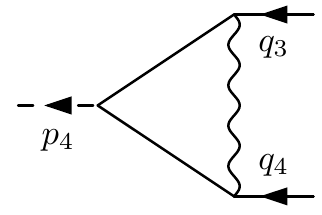} }
\right)=\nn
&
\frac{1}{\sqrt{(q-p_1-p_2)^2+((q+p_3)^2-m_h^2)^2-2(q-p_1-p_2)^2(q+p_3)^2+m_h^2)}}.
\end{align}
Therefore, the maximal cut of $\I_1$ can be written as
\begin{align}
\Cut\left(\adjustbox{valign=m}{\includegraphics[scale=0.8,trim=0 0 0.1cm -0.1cm]{fig/fig2LBox.pdf} }\right)=&
\int \D^4 k\, \delta(k^2-m^2) \delta((k-p_1)^2-m^2)\delta((k-p_1-p_2)^2-m^2)\nn\
&\times \frac{1}{\sqrt{m^4+((k+p_3)^2-m_h^2)^2-2m^2((k+p_3)^2+m_h^2)}}\,,
\label{eq:cutboxm}
\end{align}
where we have used that, when the remaining  propagators are cut, $(k-p_1-p_2)^2 = m^2$. \\
Before discussing the evaluation of the last integral, we would like to make a technical remark. When the number of on-shell conditions to be imposed is sufficiently large, the kinematic restrictions of Minkowski space could suggest that no non-trivial solution of the maximal cut exists. Therefore, one needs to adopt a practical prescription in order to obtain a result in some kinematic region and then continue it to the region of interest through a consistent relaxation of the assumptions used in the intermediate steps of the calculation. For this specific case, we found that an effective prescription consists in assuming a negative internal mass, $m^2\leq0$ .\\
Under this assumption, we evaluate the integral over the loop momentum $k$ in the rest frame of the massive external particle $p_4^\mu =(m_h, \vec{0})\,,$ by parametrizing the massless momenta $p_1^{\mu}$ and $p_2^{\mu}$ as
\begin{align}
p_{1}^{\mu}=&(E_1,0,0,E_1),\qquad p_{2}^{\mu}=(E_2,E_2\sin{\theta_{12}},0,E_2\cos{\theta_{12}}).
\label{eq:p1p2par}
\end{align}
The energies of the two particles and the relative angle between their three-dimensional momenta are expressed in terms of the Mandelstam invariants as
\begin{align}
E_{1}= \frac{s+t}{2m_h}\,, \qquad
E_{2}= \frac{m_h^2-t}{2m_h}\,,\qquad
\cos{\theta_{12}}= \frac{m_h^2(t-s)-t(s+t)}{(s+t)(m_h^2-t)}.
\label{eq:kin}
\end{align}
If we go to polar coordinates and we decompose the loop momentum as
\begin{align}
k^{\mu}=(k_0,\bar k\sin\theta\cos\varphi,\bar k\sin\theta\sin\varphi,\bar k\cos\theta),
\label{eq:kpar}
\end{align}
where we have again defined $|\vec{k}|=\bar{k}$,  we can easily check that the three additional cut-conditions  imply
\begin{align}
(k+p_3)^2=m^2-s+2m_hk_0.
\end{align}
Hence, Eq.\eqref{eq:cutboxm} becomes
\begin{align}
\Cut\left(\adjustbox{valign=m}{\includegraphics[scale=0.8,trim=0 0 0.1cm -0.1cm]{fig/fig2LBox.pdf} }\right)=&
 \int_{-\infty}^{\infty}d k_0\int_{0}^{\infty} d\bar k \bar k^2\int_{-1}^{1}d\cos\theta\int_{0}^{2\pi}d\varphi\nn
\times &\frac{\delta(k^2-m^2)\delta(2k\cdot p_1)\delta(2k\cdot p_2-s)}{\sqrt{m^4+(m^2+2m_hk_0-s-m_h^2)^2-2m^2(m^2+2m_hk_0-s+m_h^2)}}.
\end{align}
The $\delta$-function $\delta(k^2-m^2)$ can be then used to perform the integral over $\bar k$ by fixing its value to $\bar{k}=\sqrt{k_0^2-m^2}$. In this way, the constraint imposed by $\delta(2k\cdot p_1)$ reduces to a linear equation in $\cos\theta$, which yields to
\begin{align}
\Cut\left(\adjustbox{valign=m}{\includegraphics[scale=0.8,trim=0 0 0.1cm -0.1cm]{fig/fig2LBox.pdf} }\right)=&
 \frac{1}{E_1}\int_{-\infty}^{\infty}d k_0\int_{0}^{2\pi}d\varphi \,\delta(2k\cdot p_2-s)\nn
\times &\frac{1}{\sqrt{m^4+(m^2+2m_hk_0-s-m_h^2)^2-2m^2(m^2+2m_hk_0-s+m_h^2)}}.
\end{align}
Finally, we evaluate the integral over the angle $\varphi$ by using the constraint imposed by the last $\delta$-function, which is satisfied by
\begin{align}
\varphi_{\pm}=\pm \arccos{\left(\frac{2E_2(1-\cos\theta_{12})k_0-s}{2E_2\sin\theta_{12}\sqrt{-m^2}}\right)}.
\end{align}
We observe that, in order for these solutions to range in $0\leq \varphi_{\pm}\leq 2\pi$, we must restrict the integration over $k_0$ within the region
\begin{align}
k_{0,1}< k_0\leq k_{0,2},
\end{align}
with
\begin{align}
k_{0,1}=\frac{1}{2m_h}\left(s+t-2\sqrt{\frac{-t\,u\,m^2}{s}}\right)\,,\qquad k_{0,2}=\frac{1}{2m_h}\left(s+t+2\sqrt{\frac{-t\,u\,m^2}{s}}\right)\,.
\end{align}
In this way, we obtain a one-fold integral representation of the maximal cut of $\I_1$ of the type
\begin{align}
\Cut\left(\adjustbox{valign=m}{\includegraphics[scale=0.8,trim=0 0 0.1cm -0.1cm]{fig/fig2LBox.pdf} }\right)=&
 \frac{1}{ s m_h}\int_{k_{0,1}}^{k_{0,2}}d k_0\frac{1}{\sqrt{(k_{0}-k_{0,1})(k_{0}-k_{0,2})(k_{0}-k_{0,3})(k_{0}-k_{0,4})}}, 
\label{eq:rootint}
\end{align}
where we have introduced two additional roots defined as
\begin{align}
k_{0,3}=&\frac{s+m_h^2-2\sqrt{m_h^2m^2}}{2m_h}\,,\qquad k_{0,4}=\frac{s+m_h^2+2\sqrt{m_h^2m^2}}{2m_h}\,.
\end{align}
The integral \eqref{eq:rootint} can be cast into the canonical form of a complete elliptic integral of the first kind through the standard change of variables
\begin{align}
z^2=\frac{(k_0-k_{0,1}) (k_{0,4}-k_{0,2})}{(k_{0,4}-k_0) (k_{0,2}-k_{0,1})},
\end{align}
which yields to a solution of the second order differential equation \eqref{eq:2ndordm2} of the form
\begin{align}
 F_1=\Cut\left(\I_1\right)=\frac{\EK(\omega)}{ \sqrt{s^2 \left(m_h^2-t\right)^2-4 m^2 s 
 \left(m_h^2 (s-t)+t (s+t)-2 \sqrt{-s \,t \,u\,m_h^2}\right)}} \,, \label{eq:F1box}
   \end{align}
 with
 \begin{align}
 \omega=\frac{16 m^2 \sqrt{-s \,t\, u\, m_h^2}}{s
   \left(m_h^2-t\right)^2-4m^2 \left(m_h^2 (s-t)+t (s+t)-2 \sqrt{-s\,t\, u\,m_h^2}\right)}.
   \label{eq:omegabox}
 \end{align}
We observe that $F_1$ has a smooth behavior in the limit of vanishing internal mass $m^2\to 0$,
 \begin{align}
F_1 \underset{m^2\to 0}{=}\frac{1}{s(m_h^2-t)},
\end{align}
which reproduces the correct result for the maximal cut of a six-denominator box with massless propagators. Finally, using the properties of elliptic integrals, it is simple to find 
a second independent solution of Eq.~\eqref{eq:F1box} 
by changing the argument $\omega$ of the elliptic function to $1-\omega$,
\begin{align}
F_2=\frac{\EK(1-\omega)}{ \sqrt{s^2 \left(m_h^2-t\right)^2-4 m^2 s \left(m_h^2 (s-t)+t (s+t)
-2 \sqrt{-s \,t\,u\, m_h^2}\right)}}.      
\label{eq:F2box}
\end{align}
We verified explicitly that Eq.~(\ref{eq:F1box},~\ref{eq:F2box}) satisfy 
the second order differential equation~\eqref{eq:2ndordm2}, 
and of course also the corresponding ones in the other 
Mandelstam invariants $s,t$ and $m_h^2$. 
\\
We can obtain an alternative (but equivalent) representation of the solutions as follows.
We recall that the complete elliptic integral of the first kind is defined as
\begin{equation}
\EK(x) = \int_0^1 \frac{dt}{\sqrt{(1-t^2)(1-\,x\,t^2)}} \,, \quad 
\mbox{for} \quad x \in \mathbb{C} \quad \mbox{and}\quad \Re{(x)} < 1\,, 
\end{equation}
and it fulfills a second order differential equation in the form
\begin{equation}
\frac{d^2}{dx^2} \EK(x) + \left( \frac{1}{x} - \frac{1}{1-x} \right) \frac{d}{dx} \EK(x)
-\frac{1}{4}\left( \frac{1}{x} + \frac{1}{1-x} \right) \EK(x)  = 0 \,. \label{eq:deqEK}
\end{equation}
Using the transformation properties of Eq.~\eqref{eq:deqEK} under $x \to 1/x$, 
one can show that if 
$\EK(x)$ and $\EK(1-x)$ are solutions of the equation, then two other, equivalent solutions, 
are given by
$1/\sqrt{x} \EK(1/x)$ and $1/\sqrt{x} \EK(1-1/x)$.
Since any second order differential equations admits only two independent homogeneous
solutions, one must have that

\begin{equation}
\frac{1}{\sqrt{x}} \EK\left( \frac{1}{x} \right) = c_1\, \EK(x) + c_2\, \EK(1-x)\,, \qquad
\frac{1}{\sqrt{x}} \EK\left(1-\frac{1}{x} \right) = c_3\, \EK(x) + c_3\, \EK(1-x)\, \label{eq:newsol}
\end{equation}
where $c_i$, $i=1,2,3,4$ are numerical (possibly complex) constants.
Of course, since $\EK(x)$ develops a branch cut when $x>1$, one should assign a
small imaginary part to $x$, which determines the sign of the imaginary parts of the coefficients
$c_i$. 
We find, for example, for $0<x<1$ and $x \to x + i\, \delta$, that the following relations are
satisfied
\begin{equation}
\frac{1}{\sqrt{x}} \EK\left( \frac{1}{x} \right) =  \EK(x) - i\, \EK(1-x)\,, \qquad
\frac{1}{\sqrt{x}} \EK\left(1-\frac{1}{x} \right) =  \EK(1-x)\,.
\end{equation}
This means that we can equally well use either the $F_1$ and $F_2$ defined above or
the two new solutions
\begin{align}
 \widetilde{F}_1 =\frac{1}{ \sqrt{s\,m^2}  (-s\,t\,u\, m_h^2)^{1/4}} 
 \EK\left( \frac{1}{\omega} \right) \,, \qquad
  \widetilde{F}_2 =\frac{1}{ \sqrt{s\,m^2}  (-s\,t\,u\, m_h^2)^{1/4}} 
 \EK\left(1- \frac{1}{\omega} \right)\,.
  \label{eq:F1boxv2}
\end{align}
\\

Finally, it is interesting to compare our result to those of~\cite{Bonciani:2016qxi}.
There, the solution was found suitably redefining the Mandelstam variables in terms
of an additional dimensionless parameter $\alpha$, with respect to which the solution of
the equations became straightforward.
For $\alpha=1$ one should recover the standard solution.
Indeed for $\alpha=1$, the argument of the elliptic integrals found in~\cite{Bonciani:2016qxi}
reduces precisely to $1/\omega$, as in Eqs.~\eqref{eq:F1boxv2}.

%%%%%%%%%%%%%%%%%
%%%%%%%%%%%%%%%%%
\section{Conclusions and Outlook}
\label{sec:conclusions} \setcounter{equation}{0} 
\numberwithin{equation}{section} 

Differential equations are an invaluable tool for computing multi-loop
Feynman integrals. 
For any practical purposes, 
the complexity of a system of differential equations is largely 
dictated by the number of coupled differential equations that must
be solved in the limit $d \to 4$. 
First of all, one needs to find a
complete set of homogeneous solutions; once they are known,
Euler's method of the variation of constants can be used to reconstruct the
complete inhomogeneous solution.
As a matter of fact, as soon as we are left with
a system of two or more coupled irreducible differential equations, there exists 
in general no algorithm to find a complete set of homogeneous solutions;
the possibility of solving the system depends therefore on a case-by-case
analysis.

Building upon ideas developed in the context of the study of the differential equations
satisfied by the two-loop massive sunrise graph~\cite{Laporta:2004rb}, 
we showed that the maximal cut
of any given (irreducible) Feynman integral provides us precisely with one set of
homogeneous solutions of the differential equations satisfied by the latter.
This one-to-one relation can of course be inverted and, in some case, one might want to
use the solution of the homogeneous differential equations as tool to compute the maximal
cut of a graph. This is particularly useful at one-loop, since every one-loop integral 
satisfies at most a first order differential equation, whose homogeneous part
can always be solved by quadrature. In this sense, there is no need to compute
the maximal cut of any one-loop integral by direct integration over the loop momenta,
as the latter can always be obtained by solving its homogeneous differential equation,
for any arbitrary values of the dimensions $d$.

For higher loops this is not true and we are left in many cases with higher order equations.
In this case, it turns out to be much easier to compute the maximal cut 
then solving the homogeneous equation directly.
The maximal cut provides us with one set of homogeneous solutions.
When dealing with second order differential equations, 
the second solution can then be obtained in closed analytic form with standard methods.
Far from being a simple curiosity, we showed explicitly that this observation
can be turned into a very powerful tool for solving second (and possibly higher)
order differential equations.
In different non-trivial  two-loop examples, in fact, we showed how
the maximal cut can be very easily computed in 
$d=4$ space-time dimensions, providing therefore a formidable piece of information
towards the solution of coupled systems of differential equations.
A crucial step in this direction is recognizing that the calculation of the 
maximal cut of a multi-loop integral can be suitably split in the calculation of the maximal
cuts of the relevant sub-loops. 
A wise choice in the order in which sub-loops are 
cut can simplify substantially the calculation of the maximal cut.\\

Finally, we believe the results of this paper are interesting also in view of the extension
of the concept of canonical basis to Feynman integrals that do not evaluate
to iterated integrals over d-log forms only. 
Suppose, in fact, to have a $N \times N$ system of differential equations in the form
\begin{align}
\frac{\partial}{\partial x} \vec{m}(\epsilon; x) = A_0(x)\, \vec{m}(\epsilon; x) + \mathcal{O}(\epsilon)\,.
\end{align}
The maximal cut of the integrals $\vec{m}(\epsilon;x)$, computed in $\epsilon=0$ provides a 
first solution $\vec{S}^{(1)}(s)$ of the system
\begin{align}
\frac{\partial}{\partial x} \vec{S}^{(1)}( x) = A_0(x)\, \vec{S}_1(x)\,.
\end{align}
If one can use this information to find the remaining $N-1$ solutions, 
then one is left with a matrix of solutions
$$G(x) = \left( \begin{array}{ccc} S^{(1)}_1(x) & ... & S^{(N)}_1(x) \\
... & ... & ... \\
S^{(1)}_N(x) & ... & S^{(N)}_N(x) \end{array} \right)\,.$$
Rotating the master integrals to the new basis $\vec{f}(\epsilon;x)$
defined as
$$\vec{m}(\epsilon; x) = G(x)\, \vec{f}(\epsilon;x) $$
one finds by construction
\begin{align}
\frac{\partial}{\partial x} \vec{f}(\epsilon; x) = \mathcal{O}(\epsilon)\,,
\end{align}
which is one of the fundamental properties a canonical basis should fulfill.
A thorough investigation of these aspects is beyond the aim of this paper and 
will be subject of study in the near future.

\section*{Acknowledgements}
We acknowledge clarifying discussions with Kirill Melnikov, Ettore Remiddi 
and, in particular, with Pierpaolo Mastrolia. 
We also wish to thank Thomas Gehrmann, Johannes Henn and Vladimir Smirnov
for useful comments on the draft.
A.P. wishes to thank the Institute for Theoretical Particle Physics of the Karlsruhe Institute of Technology for hospitality during the development of this project.

\appendix

\bibliographystyle{bibliostyle}   
\bibliography{Biblio}
%%%%%%%%%%%%%%%%%%%%%%%%%%%%%%%%%%%%%%%%%%%%%%%%%%%%%%%%%%%%%%%%%%%%%%% 
\end{document}